\documentclass[twocolumn,showkeys,preprintnumbers,amsmath,amssymb,floatfix]{revtex4}
\usepackage{graphicx}% Include figure files
\usepackage{multirow}
\usepackage{dcolumn}% Align table columns on decimal point
\usepackage{bm}% bold math
\usepackage{color}

\newcommand{\beq}{\begin{equation}}
\newcommand{\eeq}{\end{equation}}
\newcommand{\bey}{\begin{eqnarray}}
\newcommand{\eey}{\end{eqnarray}}

\begin{document}
\preprint{}
\title{A new model for dark matter fluid sphere}
 \author{Shyam Das}
  \email{dasshyam321@gmail.com}
\affiliation{Department of Physics, P. D. Women's College, Jalpaiguri 735101, West Bengal, India}
\author{Nayan Sarkar}
 \email{nayan.mathju@gmail.com}
 \affiliation{Department of Mathematics, Jadavpur University, Kolkata 700032, West Bengal, India}
 \affiliation{Department of Mathematics, Karimpur Pannadevi College, Karimpur-741152, Nadia, West Bengal, India}

\author{Monimala Mondal}
\email{monimala.mondal88@gmail.com}
 %\affiliation{Department of Mathematics, Jadavpur University, Kolkata 700032, West Bengal, India }
\author{Farook Rahaman}
\email{rahaman@associates.iucaa.in}
 \affiliation{Department of Mathematics, Jadavpur University, Kolkata 700032, West Bengal, India }
\date{\today}% It is always \today, today,
\begin{abstract}
We develop a new model for a spherically symmetric dark matter fluid sphere containing two regions: {\bf (i)} Isotropic inner region with constant density and {\bf (ii)} Anisotropic outer region. We solve the system of field equation by assuming a particular density profile along with a linear equation of state. The obtained solutions are well-behaved and physically acceptable which represent equilibrium and stable matter configuration by satisfying the TOV equation and causality condition, condition on adiabatic index, Harrison-Zeldovich-Novikov criterion, respectively.   We consider the compact star EXO 1785-248 (Mass $M=1.3~M_{\odot}$ and radius $R = 8.8 ~km.$) to analyze our solutions by graphical demonstrations.
\end{abstract}
%Use showkeys class option if keyword
\maketitle
%\tableofcontents
\section{\label{sec1} Introduction}
One of the most interesting facts is that the ordinary baryonic matter is not the dominant form of material in the universe as is against our common perception. It is accounted that the visible component of the total mass of the universe represents only a small fraction. On the large scale structure of the cosmos, the universe is composed of matter that is fundamentally different from our familiar matter composition. Based on the astronomical observations over the years, it is accepted that most of the mass appears to be in some as-yet-undiscovered strange new form which is non-luminous known as Dark Matter (DM). The most long-lasting challenges of modern astrophysics are understanding and unveiling the nature of dark matter. It is believed that the dark matter is the main component, responsible for the large-scale structure formation in the universe. Though the dark matter is not detected in the laboratory yet still a series of observations, find the existence of dark matter in strong footing. The nature of dark matter and the standard LCDM models can be tested on many scales, from the shape of the cosmic web to the properties of galaxies and galaxy clusters, down to particle physics experiments aiming at detecting the dark matter particles. A complete understanding of the nature of dark matter component of the universe remains a mystery.

An undetected form of missing mass that emits no light, register its presence because we observe the effects of its gravity. It is found that more than $90$ percent of the mass in the whole universe is dark. One can raise the question about the understanding of the gravity which gave birth of the mysterious dark matter, but it is needless to say that the gravity is so well tested that scientist prefers the existence of dark matter.

Dark matter at the core region of the galaxies is believed to exist since long ago from the observational pieces of evidence on the stellar motion. The first observational evidence indicating the existence of the dark matter came from the work of Oort\cite{Oort} in 1932. He calculated the velocity scattered of stars in the galactic plane and found that it is greater than expected from gravitational potential from the stars. He naturally predicted more mass is needed to match the discrepancy in velocity which is apparently missing. Fritz Zwicky\cite{Zwicky} in 1933  estimated the velocity dispersion of galaxies in the Coma cluster and inferred its total mass from the Virial theorem. He found that their velocities are much beyond the velocities that could be attributed to the luminous parts of galaxies. The dynamical mass inferred from relative motions of the galaxies which found much less of the accounted mass led Zwicky to concluded the galaxies were embedded in dark halos which do not radiate.

In 1970 Vera Rubin\cite{Rubin} discovered that the rotational curves of galaxies are almost flat at large distances, which should drop as $r^{-\frac{1}{2}}$ from the galactic centre if we consider only the star density in the galaxy. The constant rotational velocity can be accounted only if one considers each galaxy must be surrounded by a super-massive halo of matter and the growth of mass with distance as one move away from the central region.  This large portion of the non-luminous form of the matter is due to dark matter. Milgrom\cite{Mond} proposed Modified Newtonian dynamics (MOND) by modifying Newton's gravity that explained the flat nature of the observed velocity rotation curve of galaxies, thus eliminating the need for hidden mass hypothesis. Later on, several observational pieces of evidence suggest that MOND cannot eliminate the need for dark matter in all astrophysical systems thus dark matter to be ubiquitous in the universe.

The presence of an extensive dark matter halo turns out to be true for all types of galaxies in general: as for an example dwarf galaxies\cite{Bell}, Stierwalt et al.\cite{Stierwalt}, many other elliptical galaxies\cite{Fabian} is a kind of common feature of the existence of DM within galaxies. Later on, the observations Roberts and Whitehurst\cite{Roberts}, Einasto et al.\cite{Einasto}, Ostriker et al.\cite{Ostriker} on different galaxies gave similar results.

Gravitational lensing is an alternate method of measuring the mass of the cluster without considering the motion of the cluster has an important contribution to establishing the existence of unseen matter. It was Zwicky\cite{Zwicky1} who suggested that gravitational lens effects could also, be used to measure the total masses of extragalactic objects. It was confirmed that the dark matter component should be there to explain the gravitational lens effects\cite{Fort}.

The best evidence to date in support of the existence of dark the matter is Bullet cluster ($1E0657-558$ )\cite{bullet} that shows a separation of ordinary matter (gas) from dark matter. Cosmic microwave background radiation properties can be used to infer the total amount of matter created by dark-matter clustering. As a cosmological evidence, Big Bang Nucleosynthesis (BBN) or as  Recent evidence hailed as the “smoking-gun” for dark matter.

Recent studies on the properties of a dark matter component declare that the dark matter particle could be Hot Dark Matter (HDM) as well as Cold Dark Matter (CDM) depending on whether the particle interaction at the initial stages of the Galaxy formation was relativistic or non-relativistic. Many exotic particles that are being proposed as a suitable candidate for dark matter including massive neutrinos, massive compact halo objects (MACHO’s) which include brown dwarfs, old white dwarfs that have ceased glowing,  neutron stars, black holes,  weakly interacting massive particles (WIMP’s)\cite{Overduin}. MACHOs do not cause enough lensing events to explain all the dark matter. Matter consisting of undiscovered particles WIMPs are excellent dark matter candidates, consistent with astronomical observations. A massless invisible and intangible neutrino can be the candidate for dark matter, but if most of the dark matter were neutrinos, they would not stay put long enough to let those structures form. Even ordinary celestial bodies such as Jupiter like objects can be supposed as a dark matter candidate. Massless scalar field that may be Brans-Dicke  type\cite{Fay} or coupled to a potential\cite{Mato} have been proposed for the dark matter component. Several attempts have been made for primordial black holes to provide an explanation for the non-baryonic dark matter.

The dark matter candidate should obey at least the law of gravity and whose mass maybe $10^{11}$ eV (weak force scale) if thermally produced in the Big Bang. Maybe as low as $10^{-5}$ eV (axion) or as high as $10^{19}$ eV (WIMPZILLA) if not thermally produced. Other plausible candidates are axions, sterile neutrinos, Supersymmetric particles, high-resolution numerical simulation favored the standard cold dark matter (SCDM) as the candidate\cite{Efstathiou,Pope}, Mirror matter, $\Lambda$-CDM\cite{Tegmark1,Tegmark2}  etc.

Dark matter is a property of gravitational effect in the aspect of the spatially geometrical structure. Earlier, it was found in the literature\cite{Bharadwaj, Su Chen} that dark matter could be described by a fluid with non-zero effective pressure. Rahaman et al.\cite{Rahaman5} considered dark matter as perfect fluid in their work. Dark matter may be modeled as a mixture of two non-interacting perfect fluids as was shown by Harko and Lobo\cite{Lobo}. Interestingly they have shown that the two-fluid model can be described as an effective single anisotropic fluid having with non-vanishing radial and transverse components of pressures\cite{Letelier,Letelier1}. The differences of pressures along radial and transverse directions are the anisotropy of a system. A review of the origins and effects of local anisotropy in astrophysical objects may also be found in Ref.\cite{Santos,Chan}.

In general, it is assumed that the galactic halo is mostly to be spherical in shape. Though the exact geometry of the galactic dark matter halo is not clear, but recent higher resolution simulations based on observational data predict the halos to be non-spherical. We shall examine the problem of modeling of dark matter in the spherically symmetric stellar configuration.

Very recent, R. P. Pant et al\cite{rp19} have studied a core-envelope model of compact star in which core is equipped with linear equation of state while the envelope is considered to be of quadratic equation of state and this work motivated us to generate a new model for stellar objects. In this work, we have proposed a new model by considering the stellar object consists of core and envelope regions. We also assume a particular EoS to describe isotropic fluid dark matter in the core region which provides constant density throughout the interior. The outer envelope region is considered as anisotropic in nature and satisfying a linear pressure-density relation. In the core boundary, we have assume de-sitter metric as the exterior while Schwarzschild solution is assumed to describe the exterior boundary of the stellar object. Accordingly, the matching conditions are used in addition to setting radial pressure zero at the exterior boundary. Energy conditions and stability has also been discussed for the developed model.

Our paper has been organized as follows: In Section~\ref{sec2}, we have presented the basic equations governing the anisotropic system. In Section~\ref{sec3}, we have presented the generalized Tolman-Oppenheimer-Volkoff (TOV) equation for anisotropic fluid distribution. By assuming a particular density profile, we have solved the relevant field equations to develop a model in Section~\ref{sec4}. In Section~\ref{sec5} the exterior metric and the corresponding boundary conditions have been displayed. We have analyzed some physical features of our model in  Section~\ref{sec6}. Finally, We have discussed and concluded our results in Section~\ref{sec7}.

\section{\label{sec2}Einstein  field equations}
We consider the line element in $Schwarzschild$ co-ordinate system to describe the interior of a static and spherically symmetric stellar configuration  as :
\begin{equation}
ds^{2}=-e^{\nu(r)}dt^{2}+e^{\lambda(r)}dr^{2}+r^{2}\left(d\theta^{2}+\sin^{2}\theta d\phi^{2} \right), \label{met}
\end{equation}
where $e^{\nu(r)}$ and $ e^{\lambda(r)} $ are known as the metric potential functions, where $\nu(r)$ and $\lambda(r)$ are functions of the radial coordinate `$r$' only.\\
 The Einstein field equations can be written as:
 \begin{equation}
 T_{\mu \nu} = \frac{1}{8\pi}\left\{R_{\mu \nu}-{1\over 2}R~g_{\mu \nu}\right\}. \label{eins}
\end{equation}
where $ T_{\mu \nu},~ R_{\mu \nu}, ~ g_{\mu \nu}$ and $R$ are  the stress energy tensor, Ricci tensor, metric tensor and Ricci scalar, respectively.\\
For an anisotropic matter distribution, the energy momentum tensor can be written as:
\begin{equation}
T_{\mu \nu} =  \{\rho(r)+p_t(r)\}U_\mu U_\nu-p_t(r) g_{\mu \nu}+\{p_r(r)-p_t(r)\}\chi_\mu \chi_\nu, \label{2}
\end{equation}
where $\rho(r)$ is the energy density, $p_r(r)$ is the radial pressure and $p_t(r)$ is the tangential pressure of the of the fluid configuration. $\chi^\mu$ is an unit $4$-vector along the radial direction and $U^\mu$ is the $4$-velocity. The quantities obey the following relation: $\chi_\mu \chi^\mu = 1$,  $\chi_\mu U^\mu = 0$. Note that we have used system of units where $G = 1 = c$.\\

The Einstein field equations (\ref{eins}) read as the following form for the metric (\ref{met}) along with the energy tensor (\ref{2}):
\begin{eqnarray}
\rho(r)&=& \frac{1}{8\pi}\left\{\frac{1-e^{-\lambda}}{r^{2}}+\frac{e^{-\lambda}\lambda'}{r}\right\},  \label{dens}\\
 p_{r}(r)&=&\frac{1}{8\pi}\left\{\frac{e^{-\lambda}-1}{r^{2}}+\frac{e^{-\lambda}\nu'}{r}\right\},  \label{prs}\\
 p_t(r)&=& \frac{e^{-\lambda}}{8\pi}\left\{\frac{\nu''}{2}+\frac{\nu'^{2}}{4}-\frac{\nu'\lambda'}{4}+\frac{\nu'-\lambda'}{2r} \right\}, \label{prt}
\end{eqnarray}
where a prime ($'$) denotes differentiation with respect to `r'.

Also, the anisotropic factor is defined  as  $\Delta(r) = \{p_t(r) - p_r(r)\}$.\\
%%%%%%%%%%%%%%%%%%%%%%%%%%%%%%%%%%%%%%%%%%%%%%%%%%%%%%%%%%%%%%%%%%%%%
\begin{figure}[!htbp]
\begin{center}
\begin{tabular}{rl}
\includegraphics[width=8cm]{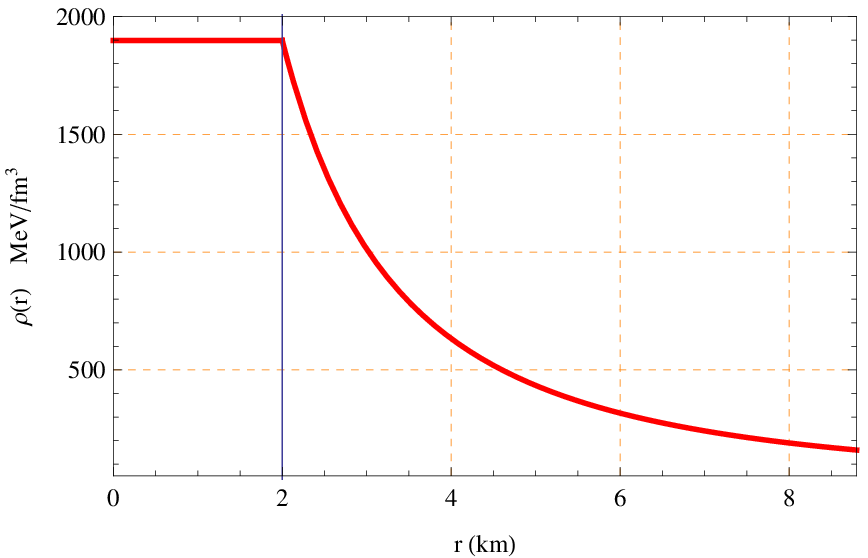}
\\
\end{tabular}
\end{center}
\caption{Behavior of the density with respect to the radial coordinate $r$ for the compact star EXO 1785-248 corresponding to the numerical value of constants given in Table-\ref{table1} }\label{fig1}
\end{figure}
%%%%%%%%%%%%%%%%%%%%%%%%%%%%%%%%%%%%%%%%%%%%%%%%%%%%%%%%%%

\section{\label{sec3}Generalized Tolman-Oppenheimer-Volkoff equation}
From Eqs.~(\ref{dens}) and (\ref{prs}), we obtain
\begin{equation}
\rho(r)+p_r(r)=\frac{\lambda^\prime+\nu^\prime}{8\pi r}e^{-\lambda},
\end{equation}
Also, from Eq.~(\ref{prs}), we get
\begin{equation}
\frac{dp_r(r)}{dr}=\frac{1}{8\pi}\left[e^{-\lambda}\left\{\frac{\nu^{\prime\prime}}{r}-\frac{\nu^{\prime}\lambda^{\prime}}{r}-\frac{\nu^{\prime}+\lambda^{\prime}}{r^2}\right\}+\frac{2(1-e^{-\lambda})}{r^3}\right],\label{dp}
\end{equation}
Then, by using Eqs.~(\ref{dens})-(\ref{dp}), we can write
\begin{equation}
-\frac{\nu'\{\rho(r)+p_r(r)\}}{2}-{dp_r(r) \over dr}-{2\{p_t(r)-p_r(r)\} \over r}=0. \label{for}
\end{equation}
The above Eq.~(\ref{for}) represents generalized Tolman-Oppenheimer-Volkoff (TOV) equation for anisotropic fluid distribution.
%%%%%%%%%%%%%%%%%%%%%%%%%%%%%%%%%%%%%%%%%%%%%%%%%%%%%%%%%%%%%%%%%%%%%
\begin{figure}[!htbp]
\begin{center}
\begin{tabular}{rl}
\includegraphics[width=8cm]{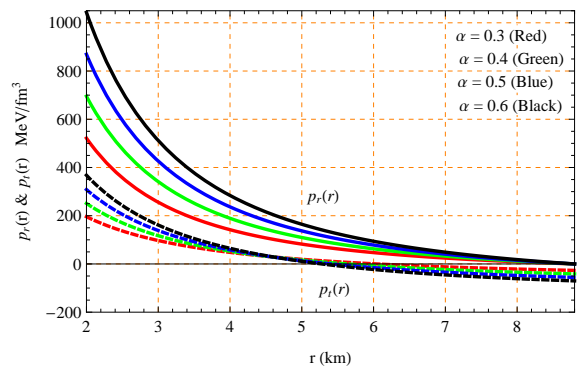}
\\
\end{tabular}
\end{center}
\caption{Behaviors of the radial and transverse pressures with respect to the radial coordinate $r$ for the compact star EXO 1785-248 corresponding to the numerical value of constants given in Table-\ref{table1}}\label{fig2}
\end{figure}
%%%%%%%%%%%%%%%%%%%%%%%%%%%%%%%%%%%%%%%%%%%%%%%%%%%%%%%%%%

\section{\label{sec4} The interior Solutions}
To solve the system of equations, we assume a density profile of the dark matter(DM) as:
\begin{equation}
\rho(r)=\frac{k}{r(1+\frac{r}{b})}\label{den},
\end{equation}
where, $b$ is the scale radius and $k$ is of dimension $km^{-1}$. We assume that the interior region of star is divided into two regions {\bf (i)} The core, $0\leq r \leq b$ and {\bf (ii)} The outer region, $b < r \leq R$ to avoid the singularity at the center of stellar configuration.

{  {Argument : Any spherically symmetric anisotropic celestial compact star is a highly dense fluid configuration with comparatively very small radius. For that reason, researchers have the curiosity to find  the exact details of the internal very dense matter compositions including its density profile. Consequently, we have motivated to find the new results which represent the highly dense fluid configuration with the above DM density profile (\ref{den}) by dividing the interior of the star in two regions as follows from the core envelope compact star model of R. P. Pant et al\cite{rp19}. Since the negative pressure is the cause of gravitational repulsion in a region which counters the inward gravitational pull and retains the region stable against the inward gravitational attraction. For this reason, we shall assume the negative pressure and positive energy density  in the core region of the compact object which remains the core stable against the inward forces acted on the core boundary. This assumption makes sure to match our interior solution with the De Sitter space as the  De Sitter space corresponds to vacuum solution of Einstein field equations with a positive energy density and negative pressure and hence with a non zero cosmological constant.}}

%%%%%%%%%%%%%%%%%%%%%%%%%%%%%%%%%%%%%%%%%%%%%%%%%%%%%%%%%%%%%%%%%%%%%
\begin{figure}[!htbp]
\begin{center}
\begin{tabular}{rl}
\includegraphics[width=8cm]{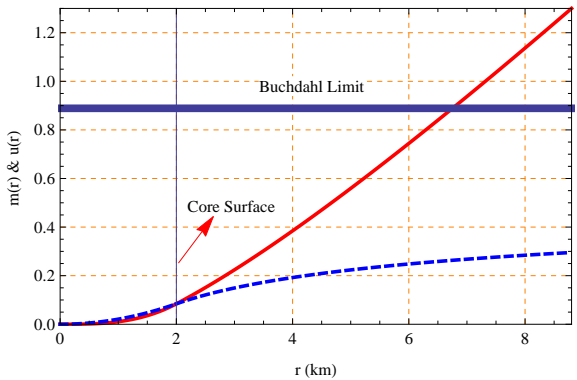}
\\
\end{tabular}
\end{center}
\caption{Behaviors of the mass and compactness parameter with respect to the radial coordinate $r$ for the compact star EXO 1785-248 corresponding to the numerical value of constants given in Table-\ref{table1} }\label{fig3}
\end{figure}
%%%%%%%%%%%%%%%%%%%%%%%%%%%%%%%%%%%%%%%%%%%%%%%%%%%%%%%%%%
\subsection{Solution in the core region,~ $0\leq r\leq b$}
We assume that the core of stellar object is isotropic in nature and satisfy the following equation of state (EoS):
\begin{equation}
p(r)=p_{r}(r) = p_{t}(r) =-\rho(r).
\end{equation}
 Negative pressure in a fluid occurs when it pushes on its surroundings. The negative pressure is the cause of gravitational repulsion around the space. This gravitational repulsion created from  pressure counters the inward gravitational pull and retain the smearing effect near the origin. \\
Therefore, from Eqs.~(\ref{for}) and (\ref{den}), we obtain
\begin{eqnarray}
\rho(r)=\text{constant}=\rho_c=\frac{k}{2b}.
\end{eqnarray}
The mass function is defined as
\begin{eqnarray}
e^{-\lambda} = 1-\frac{2m(r)}{r}.
\end{eqnarray}
Now, the mass of the core can obtain as:

\begin{eqnarray}
 m(r)&=&4\pi \int_0^r r'^2 \rho_c dr'=\frac{2r^2k\pi}{3}.
\end{eqnarray}

 Therefore, the expressions of compactness parameter and surface redshift are obtained as

\begin{equation}
 u(r)=\frac{2m(r)}{r}= \frac{4rk\pi}{3},
\end{equation}

\begin{equation}
 z(r)=\{1-u(r)\}^{-\frac{1}{2}}-1=\left(1- \frac{4rk\pi}{3}\right)^{-\frac{1}{2}}-1.
\end{equation}

In Schwarzschild coordinate, we can write
\begin{eqnarray}
e^{-\lambda} = 1-\frac{2m(r)}{r}= 1-\frac{4rk\pi}{3}.\label{e}
\end{eqnarray}

 \subsection{Solution in the outer region,~$b \leq r\leq R$}
 For the outer region, we assure the density profile (\ref{den}) and an EoS in the following linear form:
 \begin{equation}
p_r(r)=\alpha \rho-\beta,\label{eos}
\end{equation}
where $\alpha~(km^{-2}),~\beta~(km^{-2})$ are constants.\\
Now, the mass of the stellar object can  be obtained as:
 \begin{eqnarray}
 m(r)&=&4\pi \left\{\int_0^b \xi^2 \rho_c d\xi+\int_b^r \xi^2 \rho(\xi) d\xi\right\}
 \\
 &=&\frac{2bk\pi}{3}\left\{6r-5b-6bf_1\right\},
 \end{eqnarray}
 where $f_1 = \log\left(\frac{r+b}{2b}\right)$.\\
 Therefore, the compactness parameter and surface redshift  are obtained as:
 \begin{equation}
 u(r)=\frac{2m(r)}{r}= \frac{4bk\pi}{3r}\left[6r-5b-6bf_1\right],
 \end{equation}
  \begin{eqnarray}
 z(r)&=&\{1-u(r)\}^{-\frac{1}{2}}-1\nonumber
 \\
 &=&\left(1- \frac{4bk\pi}{3r}\left[6r-5b-6bf_1\right]\right)^{-\frac{1}{2}}-1.
 \end{eqnarray}
In Schwarzschild coordinate, we can write
\begin{eqnarray}
e^{-\lambda} &=& 1-\frac{2m(r)}{r},\nonumber
\\
&=& 1-\frac{4bk\pi}{3r}\left[6r-5b-6bf_1\right].\label{e}
\end{eqnarray}
On using Eqs.~(\ref{den}) and (\ref{eos}), we get  the expression of  radial pressure
\begin{eqnarray}
p_r(r) &=& \frac{\alpha k}{r(1+\frac{r}{b})}-\beta. \label{pre1}
\end{eqnarray}
On imposing Eqs.~(\ref{e})-(\ref{pre1}) in Eq.~(\ref{prs})  we get
\begin{eqnarray}
\nu'(r) &=& \frac{4\pi}{r(r+b)f_2}\Big[bk\{5b^2-br-br^2(\alpha+1)\}\nonumber
\\
&&+6r^3\beta(r+b)+6b^2k(r+b)f_1\Big],
\end{eqnarray}
whereas
\begin{equation}
f_2=\Big[4bk\pi (6r-5b)-24b^2k\pi f_1-3r\Big].
\end{equation}
%%%%%%%%%%%%%%%%%%%%%%%%%%%%%%%%%%%%%%%%%%%%%%%%%%%%%%%%%%%%%%%%%%%%%
\begin{figure}[!htbp]
\begin{center}
\begin{tabular}{rl}
\includegraphics[width=8cm]{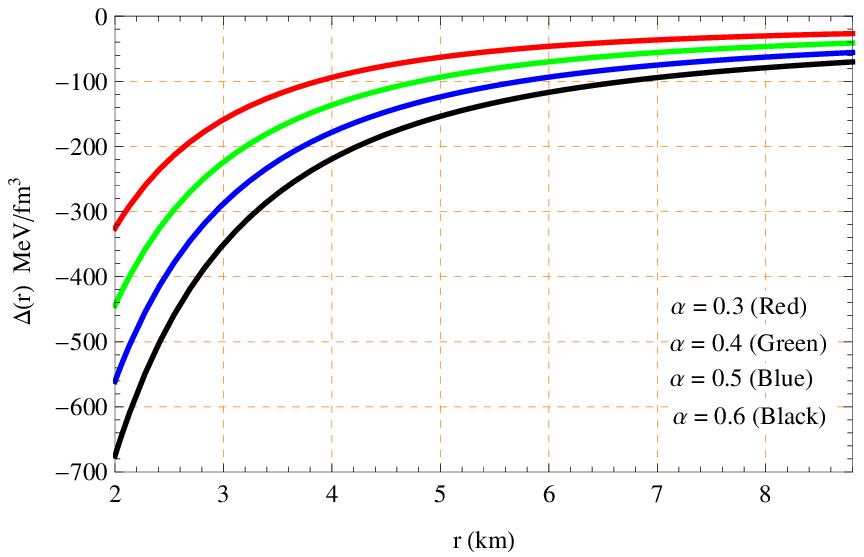}
\\
\end{tabular}
\end{center}
\caption{Behavior of the anisotropic factor with respect to the radial coordinate $r$ for the compact star EXO 1785-248 corresponding to the numerical value of constants given in Table-\ref{table1} }\label{fig4}
\end{figure}
%%%%%%%%%%%%%%%%%%%%%%%%%%%%%%%%%%%%%%%%%%%%%%%%%%%%%%%%%%
Therefore, the transverse pressure and anisotropic factor are obtained in the following exact forms:
\begin{eqnarray}
p_t(r) &=&\Big[6r\beta (r+b)f_3-12\pi r^4\beta^2(r+b)^2\nonumber
\\
&&+12\pi kb^2f_1 f_4-b^2k f_5\Big]\Big[2r(r+b)^2f_2\Big]^{-1},
\\
\Delta(r) &=&\beta-\frac{\alpha k}{r(1+\frac{r}{b})} +\Big[2r(r+b)^2f_2\Big]^{-1}\times\nonumber
\\
&&\Big[6r\beta (r+b)f_3-24\pi r^4\beta^2(r+b)^2-b^2k f_5\nonumber
\\
&&+12\pi kb^2f_1 f_4\Big],
\end{eqnarray}
whereas
\begin{eqnarray}
f_3&=&5\pi kb^3-b^2k\pi r+r^2+br\{1+4k\pi r(\alpha-1)\},\nonumber
\\
f_4&=&bk\{b(\alpha-1)-r(\alpha+1)\}-3r\beta(r+b)^2,\nonumber
\\
f_5&=&2bk\pi r(1-11\pi)+10b^2k\pi (\alpha-1)\nonumber
\\
&&+3r\{\alpha+4k\pi r (\alpha+1)^2\}.
\end{eqnarray}

\begin{figure}[!htbp]
\begin{center}
\begin{tabular}{rl}
\includegraphics[width=8cm]{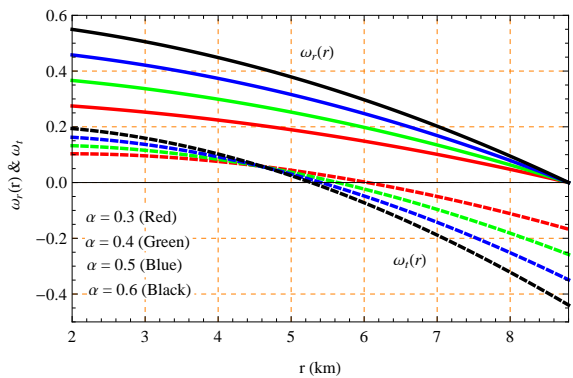}
\\
\end{tabular}
\end{center}
\caption{Behaviors of the equation of state parameters with respect to the radial coordinate $r$ for the compact star EXO 1785-248 corresponding to the numerical value of constants given in Table-\ref{table1} }\label{fig5}
\end{figure}

For observing the actual behavior of our obtained solutions, we have displayed the graphical representations of solutions in Figs. \ref{fig1}-\ref{fig5} for the well-known compact star EXO 1785-248 corresponding to $\alpha =$ 0.3, 0.4, 0.5 and 0.6. The exact behavior of the energy density is shown in  Fig. \ref{fig1}. Figs. \ref{fig2} and \ref{fig3} show the variations of radial, transverse pressures and mass with compactness parameter, respectably. Further, Figs. \ref{fig4} and \ref{fig5} display the characteristics of anisotropic factor and equation of state (EoS) parameters, respectively, where the EoS parameters are defined as $\omega_r(r) = p_r(r)/\rho(r), \omega_t(r) = p_t(r)/\rho(r)$ .

\section{\label{sec5}Boundary conditions}
 To determine the values of involved constants within solutions, we have matched our solutions with the $de$-$sitter$ metric at the core boundary $r = b$ and the exterior $Schwarzschild$ solution at the surface boundary $r = R$ $(> 2M)$.   {De Sitter space corresponds to vacuum solution of Einstein field equations with a positive energy density and negative pressure and hence with a non zero cosmological constant. In our model, the core region of the compact object we assume the pressure to be negative and energy density is positive. So it is justified to assume de Sitter metric at the core boundary. On the otherhand Schwarzschild metric is the solution Einstein field equations exterior of a spherical matter distribution with cosmological constant and charge to be zero. The compact object in our model is assumed to acquire no net charge and outer region is vacuum and hence described by Schwarzschild  solution.}\\

 The $de$-$sitter$ and exterior $Schwarzschild$ metric are given by
\begin{eqnarray}
ds_d^2&=&\left(1-\frac{r^2}{d^2}\right)dt^2-\left(1-\frac{r^2}{d^2}\right)^{-1}dr^2\nonumber
\\
&&-r^2(d\theta^2+\sin^2\theta d\phi^2),
\end{eqnarray}
\begin{eqnarray}
ds_s^2&=&\left(1-\frac{2M}{r}\right)dt^2-\left(1-\frac{2M}{r}\right)^{-1}dr^2\nonumber
\\
&&-r^2(d\theta^2+\sin^2\theta d\phi^2).
\end{eqnarray}

After matching, we have the following results:

\begin{eqnarray}
e^{-\lambda}|_{r=b} =e^{\nu}|_{r=b} = \left(1-{b^2 \over d^2}\right) =1-\frac{4\pi kb}{3},\label{bou1}
\end{eqnarray}
\begin{eqnarray}
e^{-\lambda}|_{r=R}&=&e^{\nu}|_{r=R} = \left(1-{2M \over R}\right)\nonumber
\\
&=&1-\frac{4bk\pi}{3R}\left[6R-5b-6bf_1(R)\right]. \label{bou2}
\end{eqnarray}
Again, since the radial pressure  $p_r(r)$  vanishes at the boundary $r = R$, which implies
\begin{eqnarray}
p_r(R) &=& 0. \label{bou3}
\end{eqnarray}
Using these boundary conditions (\ref{bou1})-(\ref{bou3}),  we arrived at
\begin{eqnarray}
k&=&\frac{3M}{2\pi b(6R-5b-6bf_1(R))},\nonumber
\\
d&=&\sqrt{\frac{3b}{4\pi k}},\nonumber
\\
\beta&=&\frac{\alpha bk}{R(R+b)}.
\end{eqnarray}
where $b$ is a free parameter, which will be the measurement of the core radius of the staler fluid configuration. For our model, we consider $b = 2~ km$.
%%%%%%%%%%%%%%%%%%%%%%%%%%%%%%%%%%%%%%%%%%%%%%%%%%%%%%%%%%%%%%%%%%%%%
%%%%%%%%%%%%%%%%%%%%%%%%%%%%%%%%%%%%%%%%%%%%%%%%%%%%%%%%%%

\section{\label{sec6}Physical features of the model}
Here, we are going to verify some well-established physical conditions regarding the compact star for our present solutions.

\subsection{Energy condition}
\begin{figure}[!htbp]
\begin{center}
\begin{tabular}{rl}
\includegraphics[width=8cm]{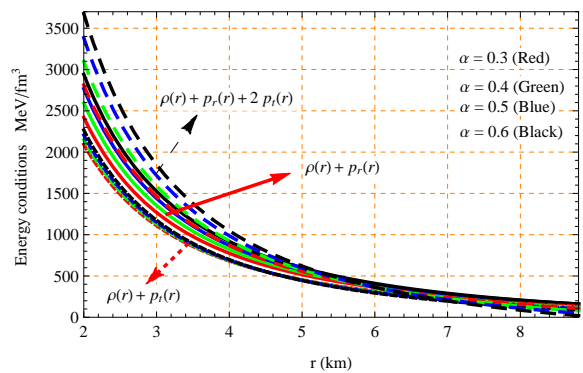}
\\
\end{tabular}
\end{center}
\caption{ Behaviors of the energy conditions with respect to the radial coordinate $r$ for the compact star EXO 1785-248 corresponding to the numerical value of constants given in Table-\ref{table1} }\label{fig6}
\end{figure}

 The energy conditions, which need to satisfy by the matter composition of stellar fluid for being physical matter\cite{Hawking73,Wald84,Maurya16}. The energy conditions are: Null energy condition (NEC), Weak energy condition (WEC) and Strong energy condition (SEC). All these energy conditions are defined in the following manner:
\begin{eqnarray}
NEC_r  &:&  \rho(r) + p_r(r)\geq 0,~~~~ NEC_t  :  \rho(r) + p_t(r)\geq 0.\nonumber
\\
WEC_r  &:& \rho(r)\geq 0,~~ \rho(r) + p_r(r)\geq 0.\nonumber
\\
WEC_t  &:& \rho(r)\geq 0,~~ \rho(r) + p_t(r)\geq 0.\nonumber
\\
SEC  &:& \rho(r) + p_r(r)+2p_t(r)\geq 0.\label{ie}
\end{eqnarray}
 Our presented solutions have satisfied all these energy conditions within outer envelop of the compact star, shown in Figs. \ref{fig1} and \ref{fig6} by plotting the LHSs of all inequalities (\ref{ie}). Moreover, the energy density is positive and $\rho(r)+p(r) = 0$ within the core region i.e. NEC and WEC are satisfied there. Consequently, the solutions represent a physical matter distribution.
%%%%%%%%%%%%%%%%%%%%%%%%%%%%%%%%%%%%%%%%%%%%%%%%%%%%%%%%%%%%%%%%%%%%%
%%%%%%%%%%%%%%%%%%%%%%%%%%%%%%%%%%%%%%%%%%%%%%%%%%%%%%%%%%

\subsection{Equilibrium condition}
\begin{figure}[!htbp]
\begin{center}
\begin{tabular}{rl}
\includegraphics[width=8cm]{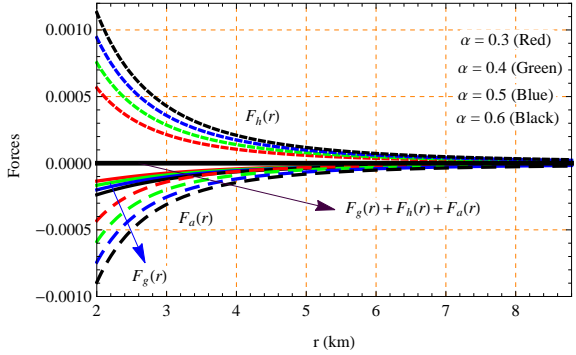}
\\
\end{tabular}
\end{center}
\caption{Behaviors of the forces with respect to the radial coordinate $r$ for the compact star EXO 1785-248 corresponding to the numerical value of constants given in Table-\ref{table1}}\label{fig7}
\end{figure}

The equilibrium position is a prime property of non-collapsing compact star. For this purpose, we are willing to verify the equilibrium condition for our solutions. Any anisotropic fluid sphere remains at equilibrium position under the action of three different forces namely, gravitational, hydrostatic and anisotropic forces, respectively. The equilibrium situation represents by an equation, which is known as the Tolman-Oppenheimer-Volkoff (TOV) equation. The generalized TOV equation for the anisotropic fluid distribution (\ref{for}) can be written as
%can be written as\cite{jr39,rct39}:
%\begin{eqnarray}
%-{M_g\{\rho(r)+p_r(r)\} \over r^2}~e^{(\lambda-\nu)/2}-{dp_r(r) \over dr}+{2\Delta(r) \over r}=0,\nonumber
%\\\label{tove}
%\end{eqnarray}
%where $M_g(r)$ is the effective gravitational mass, which can obtain  with the help of Tolman-Whittaker mass formula as:
%\begin{equation}
%M_g(r) = {r^2 \over 2} \nu' e^{(\nu-\lambda)/2}.
%\end{equation}
%Therefore, Eq.~(\ref{tove}) reduces to
%\begin{equation}
%-\frac{\nu'\{\rho(r)+p_r(r)\}}{2}-{dp_r(r) \over dr}+{2\Delta(r) \over r}=0.\label{forc}
%\end{equation}
%Eq.~(\ref{forc}) also can be written as)
\begin{equation}
F_g(r)+F_h(r)+F_a(r) = 0, \label{forc1}
\end{equation}
whereas\\
\begin{eqnarray}
\text{Gravitational force}, F_g(r)& =& -\frac{\nu'\{\rho(r)+p_r(r)\}}{2},\nonumber
\\
\text{Hydrostatic force}, F_h(r)& =& -{dp_r(r) \over dr},\nonumber
\\
\text{Anisotropic force}, F_a(r) &=& {2\{p_t(r)-p_r(r)\} \over r}.
\end{eqnarray}
The graphical demonstrations of these three forces for our solutions are shown in Fig.~\ref{fig7} and from there one can easily conclude that our model is in equilibrium state.
%%%%%%%%%%%%%%%%%%%%%%%%%%%%%%%%%%%%%%%%%%%%%%%%%%%%%%%%%%%%%%%%%%%%%
%%%%%%%%%%%%%%%%%%%%%%%%%%%%%%%%%%%%%%%%%%%%%%%%%%%%%%%%%%
\subsection{stability analysis}

\subsubsection{Causality condition}
\textbf{I. Causality condition :} Since the speed of light $c$ is the cosmic speed limit, the speed of sound within a compact star must be less than $c$ otherwise  stellar fluid is non-physical. The sound velocity inside the compact star can be determined by using
\begin{equation}
v_r(r)=\sqrt{{dp_r(r) \over d\rho(r)}},~~~v_t(r)=\sqrt{{dp_t(r) \over d\rho(r)}}.
\end{equation}
In gravitational unit, velocity of light $c = 1$. Thus $0 \leq v_r(r), v_t(r) < 1$, this condition is know as the causality condition.
%%%%%%%%%%%%%%%%%%%%%%%%%%%%%%%%%%%%%%%%%%%%%%%%%%%%%%%%%%%%%%%%%%%%%
\begin{figure}[!htbp]
\begin{center}
\begin{tabular}{rl}
\includegraphics[width=8cm]{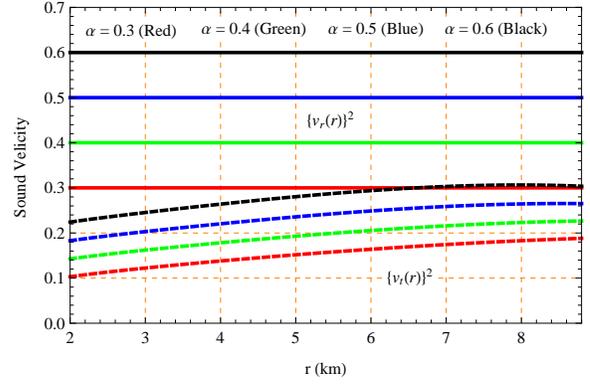}
\\
\end{tabular}
\end{center}
\caption{ Behaviors of the radial and transverse velocities of sound with respect to the radial coordinate $r$ for the compact star EXO 1785-248 corresponding to the numerical value of constants given in Table-\ref{table1} }\label{fig8}
\end{figure}
%%%%%%%%%%%%%%%%%%%%%%%%%%%%%%%%%%%%%%%%%%%%%%%%%%%%%%%%%%
Fig.~\ref{fig8} shows that our model satisfies the causality condition in outer region of the stellar object i.e. the model is of  the physical stellar fluid.\\
 \textbf{II. Stability :} To study the stability of an anisotropic fluid stellar,  L. Herrera\cite{lh92} proposed the  $cracking$ $method$  under the radial perturbations in the year 1992. Now using the concept of cracking, Abreu et al.\cite{ha07} provided the stability conditions with respect to the stability factor ($= \{v_t(r)\}^2 - \{v_r(r)\}^2$) for anisotropic fluid model. The conditions of Abreu et al.\cite{ha07}  state that {\bf:} \textbf{(i)} The region is potentially stable if $ -1 < \{v_t(r)\}^2 - \{v_r(r)\}^2 \leq 0$ and \textbf{(ii)} The region is potentially unstable if $ 0 < \{v_t(r)\}^2 - \{v_r(r)\}^2 < 1$.\\
  Our solutions satisfy the condition $ -1 < \{v_t(r)\}^2 - \{v_r(r)\}^2 < 0$ (see Fig. \ref{fig9}) within the outer region of fluid (anisotropic) configuration and therefore our model  is potentially stable.
%%%%%%%%%%%%%%%%%%%%%%%%%%%%%%%%%%%%%%%%%%%%%%%%%%%%%%%%%%%%%%%%%%%%%
\begin{figure}[!htbp]
\begin{center}
\begin{tabular}{rl}
\includegraphics[width=8cm]{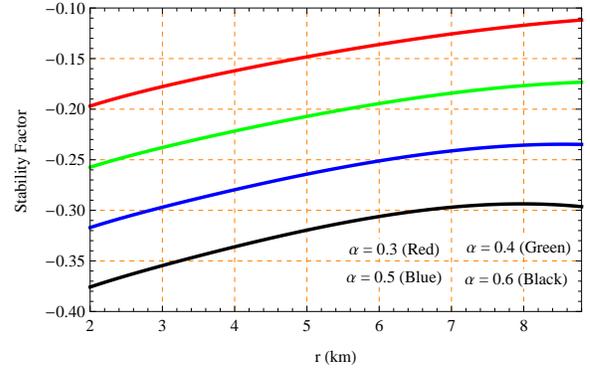}
\\
\end{tabular}
\end{center}
\caption{ Behavior of the stability factor with respect to the radial coordinate $r$ for the compact star EXO 1785-248 corresponding to the numerical value of constants given in Table-\ref{table1}}\label{fig9}
\end{figure}
%%%%%%%%%%%%%%%%%%%%%%%%%%%%%%%%%%%%%%%%%%%%%%%%%%%%%%%%%%
\subsubsection{Adiabatic index }
\begin{figure}[!htbp]
\begin{center}
\begin{tabular}{rl}
\includegraphics[width=8cm]{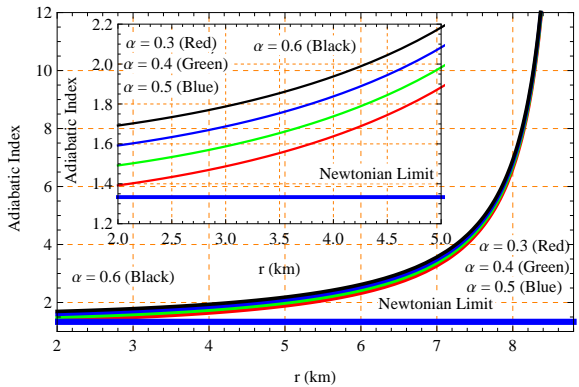}
\\
\end{tabular}
\end{center}
\caption{ Behavior of the adiabatic index with respect to the radial coordinate $r$ for the compact star EXO 1785-248 corresponding to the numerical value of constants given in Table-\ref{table1} }\label{fig10}
\end{figure}

The relativistic adiabatic index is also a considerable parameter that affect the stability of any stellar matter distribution. The relativistic adiabatic index is defined as:
\begin{eqnarray}
\Gamma_r(r) = {\rho(r)+p_r(r) \over p_r(r)}~{dp_r(r) \over d\rho(r)}
\end{eqnarray}
For the Newtonian limit, any stable configuration will alter its stability by initiating an adiabatic gravitational collapse if $\Gamma_r (r)\le 4/3$ and catastrophic if $< 4/3$ \cite{hb87}.  This condition changes for relativistic and/or anisotropic fluid which depends on the nature of anisotropy, provided by  Chan et al.\cite{rc93}.\\
 For our solutions, the value of  adiabatic index $\Gamma_r(r)$ is  more than 4/3 throughout the outer region of compact star, clear from Fig. \ref{fig10}.
%%%%%%%%%%%%%%%%%%%%%%%%%%%%%%%%%%%%%%%%%%%%%%%%%%%%%%%%%%%%%%%%%%%%%
%%%%%%%%%%%%%%%%%%%%%%%%%%%%%%%%%%%%%%%%%%%%%%%%%%%%%%%%%%
\subsubsection{Harrison-Zeldovich-Novikov criterion}
Harrison et al.\cite{har56} and Zeldovich \& Novikov\cite{zel71} proposed another important criterion for the stability of compact star. Their criterion states that the mass should increase with the increase in the central density , i.e. $dM(\rho_c)/d\rho_c > 0$. For our solutions the mass as a function of the central density can be express as
\begin{eqnarray}
M(\rho_c) &=& \frac{4\pi b^2\rho_c}{3}\left[6R-5b-6bf_1(R)\right]
\\
\frac{\partial M(\rho_c)}{\partial \rho_c} &=&  \frac{4\pi b^2}{3}\left[6R-5b-6bf_1(R)\right]>0
\end{eqnarray}
%%%%%%%%%%%%%%%%%%%%%%%%%%%%%%%%%%%%%%%%%%%%%%%%%%%%%%%%%%%%%%%%%%%%%
\begin{figure}[!htbp]
\begin{center}
\begin{tabular}{rl}
\includegraphics[width=8cm]{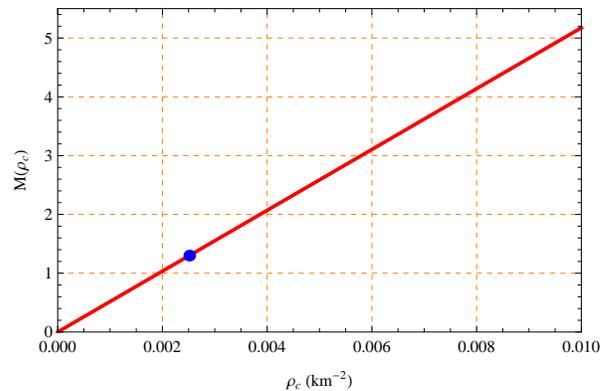}
\\
\end{tabular}
\end{center}
\caption{ Behavior of the mass with respect to the core density $\rho_c$ for the compact star EXO 1785-248 corresponding to the numerical value of constants given in Table-\ref{table1}}\label{fig11}
\end{figure}
%%%%%%%%%%%%%%%%%%%%%%%%%%%%%%%%%%%%%%%%%%%%%%%%%%%%%%%%%%
In our model, the solutions hold static stability criterion and hence stable, obvious from Fig. \ref{fig11}. Also, the dot sing in Fig. \ref{fig11} represents the corresponding mass of the compact star in solar mass unit.

\begin{table*}
 \caption{Numerical values of constants for three well-known celestial compact stars.}\label{table1}
    \begin{center}
            \begin{tabular}{ | l | l | l | l | l | l | l | l | l | }
            \hline
            \textbf{Compac Star} & \textbf{$M$}~($M_{\odot}$) & \textbf{$ R $}~($km$) & \textbf{$\alpha$}~($km^{-2}$) & $\beta$~($km^{-2}$) & $b$~ $(km)$ &~~~~$k~(km^{-1})$ & $d~ (km)$   \\ \hline
\multirow{4}{*}{EXO 1785-248}& \multirow{4}{*}{~~1.3$\pm$ 0.2}&\multirow{4}{*}{~~8.8$\pm$ 0.4}&~~~~0.3 & 0.00006    & ~~~ 2  & 0.01005 & 6.89269\\\cline{4-8}
                             &                                 &            &~~~~0.4       & 0.00008    & ~~~ 2  & 0.01005& 6.89269\\\cline{4-8}
                             &                                 &            &~~~~0.5         & 0.00011    & ~~~ 2  & 0.01005 & 6.89269\\\cline{4-8}
                             &                                 &            &~~~~0.6        & 0.00012    & ~~~ 2  & 0.01005 &6.89269\\ \hline
            \end{tabular}
            \\\vspace{0.2cm}
           \begin{tabular}{ | l | l | l | l | l | l | l | l | l | }
            \hline
            \textbf{Compac Star} & \textbf{$M$}~($M_{\odot}$) & \textbf{$ R $}~($km$) & \textbf{$\alpha$}~($km^{-2}$) & $\beta$~($km^{-2}$) & $b$~ $(km)$ &~~~~$k$~$(km^{-1})$ & d~ $(km)$   \\ \hline
\multirow{4}{*}{~~~Vela X-1}& \multirow{4}{*}{~~1.77$\pm$0.08}&\multirow{4}{*}{~~9.56$\pm$ 0.08}&~~~~0.3 & 0.00007    & ~~~ 2  & 0.012204 & 6.25493\\\cline{4-8}
                             &                                 &             &~~~~0.4 & 0.00009    & ~~~ 2  & 0.012204 & 6.25493\\\cline{4-8}
                             &                                 &             &~~~~0.5 & 0.00011    & ~~~ 2  & 0.012204 & 6.25493\\\cline{4-8}
                             &                                 &             &~~~~0.6 & 0.00013    & ~~~ 2  & 0.012204 & 6.25493\\ \hline
            \end{tabular}
            \\\vspace{0.2cm}
            \begin{tabular}{ | l | l | l | l | l | l | l | l | l | }
            \hline
            \textbf{Compac Star} & \textbf{$M$}~($M_{\odot}$) & \textbf{$ R $}~($km$) & \textbf{$\alpha$}~($km^{-2}$) & $\beta$~($km^{-2}$) & $b$~ $(km)$ &~~~~$k$~$(km^{-1})$ & d~ $(km)$   \\ \hline
\multirow{4}{*}{4U 1538-52}& \multirow{4}{*}{~~0.87}&\multirow{4}{*}{~~7.87}&~~~~0.3 & 0.000061   & ~~~ 2  & 0.00787 & 7.78763\\\cline{4-8}
                             &                                 &            &~~~~0.4 & 0.000081   & ~~~ 2  & 0.00787& 7.78763\\\cline{4-8}
                             &                                 &            &~~~~0.5 & 0.000101    & ~~~ 2  & 0.00787 & 7.78763\\\cline{4-8}
                             &                                 &            &~~~~0.6 & 0.000122    & ~~~ 2  & 0.00787 &7.78763\\ \hline
            \end{tabular}
    \end{center}
\end{table*}
\begin{table*}
\caption{Numerical values of the physical parameters for three well-known celestial compact stars corresponding to the values of constant given in Table-\ref{table1}.}\label{table2}
 \begin{center}
\begin{tabular}{| l | l | l | l | l | l | l | l |}
\hline
\textbf{Compac Star} & \textbf{$\rho_c ~(10^{15})$} & \textbf{$\rho_s ~  (10^{14})$} & \textbf{$p_{nc} ~( 10^{35})$} & $z_s$       & $u_s$      &  Buchdahl   \\
                     &  $(gm\slash cm^3)$ & $(gm\slash cm^3)$ & $(dyne\slash cm^2)$ & &       &     ~~Limit          \\\hline
 \multirow{4}{*}{EXO 1785-248}& 3.384 &  2.849 &  8.357  & 0.1914 & 0.2955 & ~~$< \frac{8}{9}$\\\cline{2-7}
                              & 3.384 & 2.849  & 11.142  & 0.1914 & 0.2955 & ~~$< \frac{8}{9}$\\\cline{2-7}
                              & 3.384 & 2.849  & 13.928  & 0.1914 & 0.2955 & ~~$< \frac{8}{9}$\\\cline{2-7}
                              & 3.384 & 2.849  & 16.714  & 0.1914 & 0.2955   & ~~$< \frac{8}{9}$\\ \hline
 \end{tabular}
 \\\vspace{0.2cm}
\begin{tabular}{| l | l | l | l | l | l | l | l |}
\hline
\textbf{Compac Star} & \textbf{$\rho_c ~(10^{15})$} & \textbf{$\rho_s ~  (10^{14})$} & \textbf{$p_{nc} ~( 10^{35})$} & $z_s$       & $u_s$      &  Buchdahl   \\
                     &  $(gm\slash cm^3)$ & $(gm\slash cm^3)$ & $(dyne\slash cm^2)$ & &       &     ~~Limit          \\\hline
 \multirow{4}{*}{~~~Vela X-1} & 4.110 & 2.975  & 10.279  & 0.2602 & 0.3703 & ~~$< \frac{8}{9}$\\\cline{2-7}
                              &4.109  & 2.975  & 13.705  & 0.2602 & 0.3703 & ~~$< \frac{8}{9}$\\\cline{2-7}
                              &4.109  & 2.975  & 17.131  & 0.2602 & 0.3703 & ~~$< \frac{8}{9}$\\\cline{2-7}
                              &4.109  & 2.975  & 20.557  & 0.2602 & 0.3703   & ~~$< \frac{8}{9}$\\ \hline
\end{tabular}
\\\vspace{0.2cm}
\begin{tabular}{| l | l | l | l | l | l | l | l |}
\hline
\textbf{Compac Star} & \textbf{$\rho_c ~(10^{15})$} & \textbf{$\rho_s ~  (10^{14})$} & \textbf{$p_{nc} ~( 10^{35})$} & $z_s$       & $u_s$      &  Buchdahl   \\
                     &  $(gm\slash cm^3)$ & $(gm\slash cm^3)$ & $(dyne\slash cm^2)$ & &       &     ~~Limit          \\\hline
 \multirow{4}{*}{4U 1538-52}& 2.651 &  2.731 &  6.412  & 0.1331 & 0.2211  & ~~$< \frac{8}{9}$\\\cline{2-7}
                              & 2.651 & 2.731  & 8.549   & 0.1331  & 0.2211 & ~~$< \frac{8}{9}$\\\cline{2-7}
                              & 2.651 & 2.731  & 10.687  & 0.1331 & 0.2211  & ~~$< \frac{8}{9}$\\\cline{2-7}
                              & 2.651 & 2.731  & 12.824  & 0.1331 & 0.2211  & ~~$< \frac{8}{9}$\\ \hline
 \end{tabular}
 \end{center}
\end{table*}

\section{\label{sec7} Discussions and conclusion }

In this paper, we have presented a model for spherically symmetric dark matter (DM) fluid (isotropic as well as anisotropic) sphere, where the DM is characterized by a density profile $\rho(r)=k/r(1+\frac{r}{b})$ . For having the singularity of the density profile a the centre of celestial compact star, we have assumed that the star is divided into two regions: {\bf A.} The isotropic core region with equation of state(EoS) $p_r(r) = -\rho(r)$ and {\bf B.} The anisotropic outer region along with EoS $p_r(r) = \alpha\rho(r)-\beta$. In the core region, the density  becomes constant depending upon the EoS, which we have assumed to avoid the singularity at the centre. To analyze our obtained solutions with the help of graphical representations, we have considered the well-known compact star EXO 1785-248. Moreover, we have calculated the numerical values of all physical parameters for the compact stars EXO 1785-248 along with more well-known compact stars Vela X-1 and 4U 1538-52 in tabular form to make our solutions more feasible.\\

The salient key features of our solutions are:\\

\textbf{(i) Energy density and pressures:} The energy density of a compact star should be positive inside the star. In our solutions, the energy density is positive with constant and monotonically decreasing in nature towards the surface within the inner region and outer region, respectively. Moreover, inner and outer values of the energy densities are coincided at the inner surface boundary $b = 2~km$, clear from Fig. \ref{fig1}. The pressures (radial and transverse) are monotonically decreasing towards the surface in outer envelope. In the inner region, the pressure (radial = transverse) becomes negative depending on EoS and due to negative pressure, the inner part pushes back the surrounding matter to remain at the same size. However, the radial pressure is positive within the outer part of the compact star and vanishes at the outer surface but the transverse pressure has changed its phase from positive to negative around $r = 6~km$ (see Fig. \ref{fig2}). Also, we have computed the numerical values of core energy density $\rho_c$ and surface (outer region) energy density $\rho_s$ for three well-known compact stars and all are of orders $10^{15}$ and $10^{14}$, respectively. The numerical values of pressures $p_{nc}$ very near to the core boundary from the side of outer region are of orders $10^{35}$ . Therefore, all these values are physically well-valued.\\

\textbf{(ii) Anisotropy:} For the isotropy, the anisotropic factor $\Delta(r) = 0$ in inner part. However, in the outer region, the radial pressure is always greater than the transverse pressure (see Fig. \ref{fig2}) in our solutions. Therefore, the anisotropy becomes negative in nature, also obvious from Fig. \ref{fig4}. This nature of anisotropy indicates that the force due to anisotropy is inward-directed i.e. compact star becomes less stable, that can be also seen from Fig. \ref{fig7}.\\

\textbf{(iii) Equation of state parameters:} The EoS $p_r(r) = -\rho(r)$ shows that the EoS parmeter $\omega(r) = -1$. For the real feasible matter distribution, the radial EoS parameter $\omega_r(r)$ should be lie in $0 < \omega_r(r) < 1$\cite{fr10}. The obtained solution has satisfied the condition $0 < \omega_r(r) < 1$, provided in Fig. \ref{fig5} in outer part of the fluid configuration and hence the DM becomes as real feasible matter within outer region. Thus, it is the beauty of our solutions regarding the celestial compact stars, which are formed by DM distributed in two parts: {\bf (i)} Inner part, formed by unfeasible DM, {\bf (ii)} Outer part, formed by real feasible DM.\\

\textbf{(iv) Energy conditions:} The satisfaction all energy conditions are necessary for physical matter distribution and evidently, we can see from Fig.\ref{fig6} along with Fig.\ref{fig1} that our solutions satisfied all the energy conditions within outer envelope of star. Moreover, inner part has maintained the NEC and WEC. Therefor, the matter distribution is physical one. \\

\textbf{(v) Equilibrium:} Under the action of gravitational, hydrostatic and anisotropic forces any celestial anisotropic fluid configuration remains at the equilibrium position. The matter distribution represented by our solutions is in the equilibrium position, clear from Fig.\ref{fig7}. In equilibrium position, hydrostatic is repulsive and   gravitational, anisotropic forces are attractive in nature.\\

\textbf{(vi) Mass function and compactness parameter: } The exact behaviors of mass function $m(r)$ and compactness parameter $u(r)$ are demonstrated  in Fig.\ref{fig3}. From that figure, we can see that  $m(r)$ and $u(r)$ are finite, zero at the centre and then monotonically increasing toward the outer layer surface of the fluid configuration. Moreover, the value of compactness parameter is more that the mass in the region $ 0 < r < b$ and coincide at the core boundary $r = b = 2~km$ (see Fig.\ref{fig3}).  According to Buchdahl\cite{ha59}, the value of the compactness parameter at the outer layer surface of compact star, $u_s = u(R) = \frac{2M}{R} < \frac{8}{9}$. We have calculated the numerical values $u_s$, provided in Table-\ref{table2} and all these values ensure that our solutions satisfied the Buchdahl limit.\\

\textbf{(vii) Stability: } The stable situation is one of the essential characteristics of compact star. We have analyzed the stability situation with the help of causality condition, adiabatic index and Harrison-Zeldovich-Novikov criterion. The radial and transverse velocities of sound are positive and less than 1 (see Fig.\ref{fig8}) and hence our solutions satisfied the causality condition within the outer region of compact fluid configuration. The stability factor $\{v_r(r)\}^2 - \{v_r(t)\}^2$ is negative, shown in Fig.\ref{fig9}, i.e. DM within outer region is physical matter, which are potentially stable.   The profiles of adiabatic index and mass in terms of central density also indicate that our solutions represent stable matter configuration, clear from Fig.\ref{fig10} and Fig.\ref{fig10}, respectively.\\

\textbf{(viii) Surface redshift:}  The variation of obtained surface redshift $z(r)$ is shown in Fig. \ref{fig11}. From that figure, we can see that the  $z(r)\rightarrow 0$ as $r \rightarrow 0$ and thereafter monotonically increasing unto the surface. Moreover, the inner and outer values of  $z(r)$ have coincided at the inner boundary. Further, we have computed the maximum numerical values of surface redshift $z_s$ = $z(R)$ at the outer boundary of the celestial fluid configuration, provided in Table-\ref{table2} and all these values of $z_s$ are within the range provided by the author of \cite{bv02}.\\

Finally, all the salient key features of our solutions ensure that our solutions are well-behaved and physically acceptable to represent the physical DM fluid configuration containing two parts: the isotropic inner part of unfeasible DM with constant density and anisotropic outer part filled with feasible DM.

\begin{acknowledgments}
Farook Rahaman and Shyam Das gratefully acknowledge support from the Inter-University Centre for Astronomy and Astrophysics (IUCAA), Pune, India, where part of this work was carried out. FR is grateful to DST-SERB, Govt. of India and RUSA 2.0, Jadavpur University for financial support. We are thankful to the reviewers for their valuable suggestions.
\end{acknowledgments}

\end{document}